\documentclass[aps,pre,showpacs, twocolumn]{revtex4} 
\usepackage{graphicx}
\usepackage{dcolumn}
\usepackage{amsmath}
\usepackage{amssymb}
\usepackage{color}
\usepackage{bm}

\newcommand{\bvec}[1]{\mbox{\boldmath $#1$}}

\begin{document}
\preprint{HEP/123-qed}
\title{Non-equilibrium quasi-long-range order of driven random field O($N$) model}
\author{Taiki Haga}
\affiliation{Department of Physics, Kyoto University, Kyoto 606-8502, Japan}
\email[]{haga@scphys.kyoto-u.ac.jp}
\date{\today}

\begin{abstract}
We investigate three-dimensional O($N$) spin models driven with a uniform velocity over a random field.
Within a spin-wave approximation, it is shown that in the strong driving regime the model with $N=2$ exhibits a quasi-long-range order 
in which the spatial correlation function decays in a power-law form.
Furthermore, for the cases that $N=2$ and $3$, we numerically demonstrate a non-equilibrium phase transition between the quasi-long-range order phase and the disordered phase,
which turns out to resemble the Kosterlitz-Thouless transition in the two-dimensional pure XY model in equilibrium.
\end{abstract}

\pacs{05.70.Fh, 05.60.-k, 75.10.Nr}

\maketitle

\section{Introduction}
Thermodynamic systems can exhibit remarkably complex phase transition dynamics when they are driven by an external force.
Well-studied examples include ferromagnetic systems and crystalline systems that are driven by steady shear \cite{Onuki,Miyama}, 
oscillating external fields \cite{Chakrabarti,Klapp}, and athermal noise \cite{Sancho,Behn}.
The general treatment that concerns how the driving forces affect the nature of phase transitions has not been established yet.
To identify and classify various types of phase transitions peculiar to out of equilibrium systems can give valuable insights for 
developing non-equilibrium statistical mechanics.

The ferromagnetic and crystalline phases are characterized by long-range order (LRO), wherein the order parameter takes a finite value.
As a qualitatively different behavior, the two-dimensional XY model exhibits quasi-long-range order (QLRO), wherein the order parameter remains zero but 
the spatial correlation decays in a power-law form at low temperatures \cite{KT}.
Furthermore, at the transition between the QLRO phase and the disordered phase, which is called the Kosterlitz-Thouless (KT) transition, 
there is no singularity in thermodynamic quantities such as specific heat and susceptibility \cite{Gupta}.
This peculiar type of phase transition has attracted considerable research attention.
However, the possibility that a driving force causes the KT transition has not been clearly discussed or established.
Consequently, the question arises as to whether there is a system that does not exhibit the KT transition in equilibrium, while it does exhibit the transition in the presence of a driving force.
To the best of our knowledge, such a system has not been reported thus far.
In this paper, we consider three-dimensional O($N$) spin models driven with a uniform velocity over a random field as examples of such systems.
This ``non-equilibrium KT transition'' may define a novel type of universality class, wherein the interplay between disorder and driving plays a crucial role.

The dynamics of an ordered system driven by an external force over a random substrate have been a topic of intensive research in statistical physics.
A rich variety of complex phenomena results from the interplay between elasticity, quenched disorder and driving.
The best-known example of such systems may be driven vortex lattices in dirty superconductors \cite{LeDoussal98,Stroud,Bishop}.
In such systems, impurities and crystalline defects act as random pinning potentials for vortex lines.
There are other well-studied systems, e.g., charge-density waves \cite{Gruner} or colloids \cite{Reichhardt} driven by an external field.
The driven vortex lattice systems exhibit dynamical reordering transition for a finite driving velocity.
In equilibrium, the vortex lattice system has two different phases \cite{Menon}.
In the strong disorder regime, the vortex glass phase is realized, in which the spatial correlation function of the displacements of the vortices exponentially decays.
In the weak disorder regime, the Bragg glass phase is realized, in which the spatial correlation function exhibits power-law decay \cite{LeDoussal94}.
When the vortex glass is driven with a finite velocity, the effective disorder that the vortices experience becomes weaker because the random potential varies rapidly in a moving frame.
At a large driving velocity, a first-order phase transition to the Bragg glass phase from the vortex glass phase occurs \cite{Stroud,Bishop}.
While the vortex lattice systems that exhibit the first-order phase transition in the absence of the disorder are well studied, very few studies have focused on 
a driven system over a random substrate that exhibits a second-order phase transition in the absence of disorder.

In this paper, we consider the nature of the dynamical reordering transition of three-dimensional O($N$) spin models when they are driven with a uniform velocity over a random field.
We show that the models with $N=2$ and $3$ exhibit QLRO at low temperatures and that the transition from the QLRO phase to the disordered phase resembles the KT transition.
This paper is organized as follows:
In Sec.~\ref{sec:Model}, we introduce the models and review their behavior in equilibrium.
In Sec.~\ref{sec:SW}, for the case that $N=2$ (XY model), we show that this model exhibits QLRO at low temperatures by using the spin-wave approximation.
In Sec.~\ref{sec:results}, the phase diagram for $N=2$ and $3$ with respect to temperature, disorder, and driving velocity is determined by means of numerical experiments.
We calculate the specific heat and show that it exhibits no singularity at the transition point.
In Sec.~\ref{sec:conclusions}, we summarize our results.
We also discuss nematic liquid crystals flowing in a random medium as an experimental realization of our model.

\section{Model}
\label{sec:Model}
Let $\{\phi^{\alpha} \}^N_{\alpha=1}$ be an $N$-component ($N \geq 2$) real vector field.
The Hamiltonian of the three-dimensional O($N$) models with a quenched random field is given by
\begin{eqnarray}
H[\bvec{\phi};\bvec{h}]=\int {\rm d}^3 \bvec{r} \biggl[ \frac{1}{2}K|\bvec{\nabla} \bvec{\phi} (\bvec{r})|^2-\frac{r}{2} |\bvec{\phi}(\bvec{r})|^2  \nonumber \\
+\frac{g}{4N}| \bvec{\phi}(\bvec{r})|^4- \bvec{h}(\bvec{r}) \cdot  \bvec{\phi}(\bvec{r})  \biggr] ,
\end{eqnarray}
where $K$, $r$, and $g$ represent positive parameters.
The quenched random field $\bvec{h}(\bvec{r})$ obeys a Gaussian distribution with $ \langle \bvec{h}(\bvec{r}) \rangle = \bvec{0} $ and 
$ \langle h^{\alpha}(\bvec{r}) h^{\beta}(\bvec{r'}) \rangle=h_0^2 \delta_{\alpha \beta} \delta(\bvec{r}-\bvec{r'})$.
The time evolution of the field $\bvec{\phi}(\bvec{r},t)$ is described by
\begin{equation}
\frac{\partial \phi^{\alpha}(\bvec{r},t)}{\partial t}+\bvec{v} \cdot \bvec{\nabla} \phi^{\alpha}(\bvec{r},t)=-\Gamma \frac{\delta H[\bvec{\phi};\bvec{h}]}{\delta \phi^{\alpha}(\bvec{r},t)}+\eta^{\alpha}(\bvec{r},t),
\label{EM-LF}
\end{equation}
where $\bvec{v}=v \bvec{e_x}$ denotes a parameter independent of $(\bvec{r},t)$, and $\eta^{\alpha}(\bvec{r},t)$ represents thermal noise that satisfies 
$ \langle \eta^{\alpha}(\bvec{r},t) \eta^{\beta}(\bvec{r'},t') \rangle=2 \Gamma T \delta_{\alpha \beta} \delta(\bvec{r}-\bvec{r'}) \delta(t-t')$.
This equation describes the relaxation dynamics of the ordered system that is driven with a uniform velocity over the quenched random field.
Examples of such systems include nematic liquid crystals flowing in a random medium such as an aerogel or a porous substrate, 
where the random field corresponds to the random anchoring that occurs due to the complicated surface structure of the porous substrate.
We investigate the non-equilibrium steady states of the models with $N=2$ (XY model) and $N=3$ (Heisenberg model).
We call these models as driven random field O($N$) models (DRFO($N$)Ms).

Let us review the behavior of these models in equilibrium ($\bvec{v}=0$).
The random field O($N$) models (RFO($N$)Ms) are one of the simplest disordered systems in which spins with O($N$) symmetry are coupled to a quenched random field.
The Imry-Ma argument and ``dimensional reduction'' property state that LRO with continuous symmetry breaking is destroyed by an infinitesimally weak random field below four dimensions \cite{Imry-Ma,Aharony,Parisi}.
The absence of LRO below four dimensions was also rigorously proved by Aizenman and Wehr \cite{Aizenman}.
While the theoretical description that concerns the existence of LRO has been established, whether QLRO exists or not in three dimensions is a more subtle problem.
From the analogy of the Bragg glass phase in a vortex lattice system \cite{LeDoussal94}, 
theoretical studies based on the renormalization group analysis have suggested the existence of QLRO for the three-dimensional random field XY model and the random anisotropy Heisenberg model \cite{Fisher85,Feldman}.
In order to confirm this predicted QLRO, numerical simulations have also been conducted \cite{Gingras,Itakura}.
However, definitive evidence for the existence of QLRO has not thus far been obtained because the correlation length rapidly increases when the strength of the random field decreases.
Moreover, in Ref.~\cite{Zannoni00}, the experimental investigation of nematic liquid crystals in aerogels did not lead to the observation of any QLRO.
Recently, more sophisticated renormalization group studies have negated the existence of QLRO in three dimensions \cite{Tissier,LeDoussal06}.
In summary, to the best of our knowledge, we can conclude that three-dimensional RFO($N$)Ms with $N \geq 2$ do not exhibit any phase transitions.

\section{Spin-wave approximation}
\label{sec:SW}
Let us calculate the spin correlation function for the XY model ($N=2$) by using the spin-wave approximation at zero temperature.
The magnitude of the spin is assumed to be fixed to unity. The order parameter field is represented by $ \bvec{\phi}(\bvec{r})=(\phi^1(\bvec{r}), \phi^2(\bvec{r}))=( {\rm cos}\theta(\bvec{r}), {\rm sin}\theta(\bvec{r}) ) $.
The Hamiltonian is rewritten as
\begin{equation}
H[\theta]=\int {\rm d}^3 \bvec{r} \left[ \frac{1}{2}K(\nabla \theta(\bvec{r}))^2-h(\bvec{r}){\rm cos}(\theta(\bvec{r})-\xi(\bvec{r})) \right],
\end{equation}
where the random field is written as $ \bvec{h}(\bvec{r})=(h^1(\bvec{r}),h^2(\bvec{r}))=(h(\bvec{r}) {\rm cos}\xi(\bvec{r}), h(\bvec{r}) {\rm sin}\xi(\bvec{r})) $.
The equation of motion at zero temperature is given by
\begin{eqnarray}
\frac{\partial \theta(\bvec{r},t)}{\partial t} + v \frac{\partial \theta(\bvec{r},t)}{\partial x}=\Gamma \bigl[K \nabla^2 \theta(\bvec{r},t)  \nonumber \\
-h(\bvec{r}){\rm sin}(\theta(\bvec{r},t)-\xi(\bvec{r})) \bigr].
\end{eqnarray}
We define the Green function $G(\bvec{r})$ by its Fourier transform ${\tilde G}(\bvec{q})=(\Gamma K q^2 +ivq_x )^{-1}$.
Subsequently, the formal solution for the steady state is given as
\begin{eqnarray}
\theta(\bvec{r})=\int {\rm d} \bvec{r'} G(\bvec{r}-\bvec{r'}) \nonumber \\
\times \Gamma \left\{-h^1(\bvec{r'}){\rm sin}\theta(\bvec{r'})+h^2(\bvec{r'}){\rm cos}\theta(\bvec{r'}) \right\}.
\end{eqnarray}
The mean square relative displacement $ \langle (\theta(\bvec{r_1})-\theta(\bvec{r_2}))^2 \rangle $ is calculated as
\begin{eqnarray}
\langle (\theta(\bvec{r_1})-\theta(\bvec{r_2}))^2 \rangle \nonumber \\
= \int {\rm d}^3 \bvec{r'} {\rm d}^3 \bvec{r''} \left\{G(\bvec{r_1}-\bvec{r'})-G(\bvec{r_2}-\bvec{r'}) \right\} \nonumber \\
\times \left\{G(\bvec{r_1}-\bvec{r''})-G(\bvec{r_2}-\bvec{r''}) \right\} \nonumber \\
\times \Gamma^2 \{ \: \langle h^1(\bvec{r'})h^1(\bvec{r''}){\rm sin}\theta(\bvec{r'}){\rm sin}\theta(\bvec{r''}) \rangle \nonumber \\
-2 \langle h^1(\bvec{r'})h^2(\bvec{r''}){\rm sin}\theta(\bvec{r'}){\rm cos}\theta(\bvec{r''}) \rangle \nonumber \\
+ \langle h^2(\bvec{r'})h^2(\bvec{r''}){\rm cos}\theta(\bvec{r'}){\rm cos}\theta(\bvec{r''}) \rangle \: \}.
\end{eqnarray}
We use factorization approximations such as 
$ \langle h^{\alpha}(\bvec{r})h^{\beta}(\bvec{r'}){\rm sin}\theta(\bvec{r}){\rm sin}\theta(\bvec{r'}) \rangle \simeq 
 \langle h^{\alpha}(\bvec{r})h^{\beta}(\bvec{r'}) \rangle \nonumber \\
\times \langle {\rm sin}\theta(\bvec{r}){\rm sin}\theta(\bvec{r'}) \rangle $ \cite{Garanin}.
This factorization is justified when the correlation length of $ \theta(\bvec{r}) $, which is denoted by $ \xi $, is much larger than that of the random field $ \xi_R $.
We will consider a self-consistent condition leading to $ \xi \gg \xi_R $ after the calculation of the mean square relative displacement.
If this condition is satisfied, we have
\begin{eqnarray}
\langle (\theta(\bvec{r_1})-\theta(\bvec{r_2}))^2 \rangle \nonumber \\
= \Gamma^2 h_0^2 \int {\rm d} \bvec{r'} \left\{G(\bvec{r_1}-\bvec{r'})-G(\bvec{r_2}-\bvec{r'}) \right\}^2.
\label{MSRD-G}
\end{eqnarray}
Substituting the explicit form of the Green function, we have
\begin{eqnarray}
\langle (\theta(\bvec{r_1})-\theta(\bvec{r_2}))^2 \rangle \nonumber \\
= 2 \Gamma^2 h_0^2 \int \frac{{\rm d}^3 \bvec{q}}{(2 \pi)^3} \frac{1-{\rm cos} \left\{ \bvec{q} \cdot (\bvec{r_1}-\bvec{r_2}) \right\} }{\Gamma^2 K^2 q^4+v^2 q_x^2}.
\label{MSRD}
\end{eqnarray}

From Eq.~(\ref{MSRD}), we obtain the asymptotic behavior over a large distance $r \gg \Gamma K/v$,
\begin{eqnarray}
\langle (\theta(\bvec{r})-\theta(\bvec{0}))^2 \rangle \simeq
  \begin{cases}
    \frac{\Gamma h_0^2}{4 \pi K v} {\rm ln}\: r, \:\:(\bvec{r} \parallel \bvec{v}), & \\
    \frac{\Gamma h_0^2}{2 \pi K v} {\rm ln}\: r, \:\:(\bvec{r} \perp \bvec{v}). &
  \end{cases}
  \label{MSRD-asymp}
\end{eqnarray}
The detailed calculation is presented in the Appendix A.
This logarithmic dependence on the distance is similar to that of the KT transition in the two-dimensional pure XY model \cite{Goldenfeld}.

Let us consider a self-consistent condition leading to $ \xi \gg \xi_R $.
We define the correlation length $ \xi $ by $ \langle (\theta(\xi)-\theta(0))^2 \rangle \sim 1 $.
The correlation length of the random field $ \xi_R $ is equal to a cut-off for short length scale $ \Lambda^{-1} $ 
because the correlation function is given by $ \langle h^{\alpha}(\bvec{r}) h^{\beta}(\bvec{r'}) \rangle=h_0^2 \delta_{\alpha \beta} \delta(\bvec{r}-\bvec{r'})$.
If we suppose $ \xi \gg \xi_R $ as an ansatz, we obtain from the above calculation
\begin{equation}
\frac{\xi}{\Lambda^{-1}} \sim {\rm exp} \left[ \frac{2 \pi K v}{\Gamma h_0^2} \right],
\end{equation}
which becomes infinitely large in the weak disorder limit or in the strong driving limit.
This argument suggests that $ \xi \gg \xi_R $ when
\begin{equation}
\frac{2 \pi K v}{\Gamma h_0^2} \gg 1.
\label{scale-separation}
\end{equation}
Thus, if the condition Eq.~(\ref{scale-separation}) is satisfied, 
the factorization approximation used in the calculation of Eq.~(\ref{MSRD-G}) is justified.
In the next section, we will confirm that this condition is satisfied in the region of QLRO.

The correlation function $ C(r) \equiv \langle \bvec{\phi}(\bvec{r}) \cdot \bvec{\phi}(\bvec{0}) \rangle $ is calculated as follows:
\begin{eqnarray}
C(r) = \langle e^{i(\theta(r)-\theta(0))} \rangle \nonumber \\
= {\rm exp} \left[ \sum_{n=1}^{\infty} \frac{1}{n!} i^n \langle (\theta(r)-\theta(0))^n \rangle_c \right],
\end{eqnarray}
where $ \langle (...)^n \rangle_c $ denotes a $n$-th cumulant.
If one admits the factorization approximation noted above,
it is shown that all higher cumulants approximately vanish because the random field $h^{\alpha}$ obeys a Gaussian distribution.
Thus, we have $C(r) = e^{-1/2 \langle (\theta(r)-\theta(0))^2 \rangle} $.
From Eq.~(\ref{MSRD-asymp}), we have
\begin{eqnarray}
 C(r) \propto
  \begin{cases}
    r^{-\alpha_{\parallel}}, \:\:(\bvec{r} \parallel \bvec{v}), & \\
    r^{-\alpha_{\perp}}, \:\:(\bvec{r} \perp \bvec{v}), &
  \end{cases}
\end{eqnarray}
where the exponents are $\alpha_{\parallel}=\Gamma h_0^2/(8 \pi Kv)$ and $\alpha_{\perp}=\Gamma h_0^2/(4 \pi Kv)$.
Therefore, we have shown that DRFO(2)M exhibits anisotropic QLRO at low temperatures, in which the spin-wave approximation is valid.

\section{Numerical results}
\label{sec:results}
We investigate the transition between the QLRO phase and the disordered phase by numerically solving Eq.~(\ref{EM-LF}).
The calculation is implemented in a moving frame with velocity $\bvec{v}$.
The periodic boundary conditions and free boundary conditions are imposed for the directions perpendicular and parallel to the driving velocity $\bvec{v}$, respectively.
The random field is continuously generated in the front boundary of the simulation box, and it moves with velocity $-\bvec{v}$.
The detailed method of the numerical calculation is presented in the Appendix B.
Time integration is performed by employing the Euler method.
The parameter values are fixed as $K=1$, $\Gamma=1$, $r=5$, and $g=10$.
We set the time and space discretization as $ \delta t=0.005 $ and $ \delta x=1 $, respectively.

\subsection{XY model}

\begin{figure}
 \centering
 \includegraphics[width=60mm]{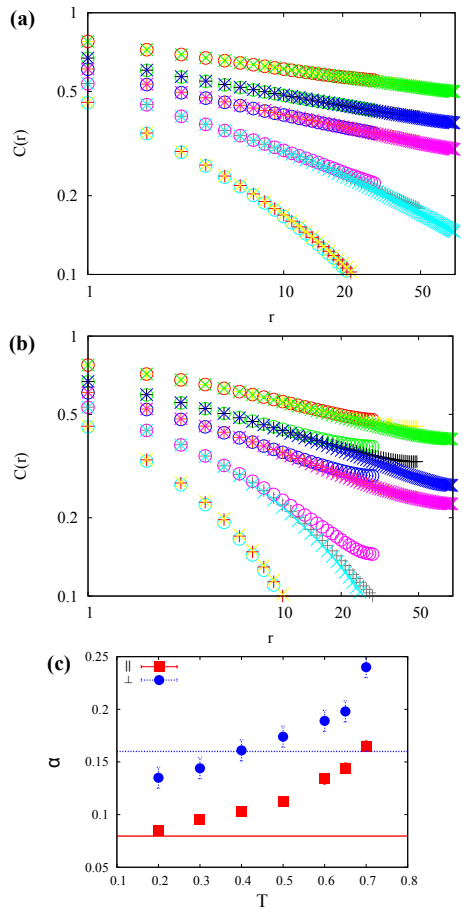}
 \caption{(Color online) Spin correlation function $C(r)$ for different values of temperature.
 The upper (a) and lower (b) panels depict $C(r)$ for the directions parallel and perpendicular to the driving velocity $\bvec{v}$, respectively.
 The values of temperature are $T=0.40$, $0.60$, $0.70$, $0.80$ and $0.90$ from the top to the bottom of the panels.
 The other relevant parameters are $h_0=1.0$ and $v=0.5$.
 The transition temperature is estimated as $T_{\rm c}=0.70 \pm 0.05$.
 The symbols $ \circ $, $ + $, and $ \times $ represent $C(r)$ for system sizes of $60^3$, $100^3$, and $150^3$, respectively.
 The error-bars are comparable with the size of the symbols.
 The panel (c) shows the temperature dependence of the exponents $\alpha_{\parallel}$ and $\alpha_{\perp}$ determined from the data corresponding to the system size of $60^3$.
 The error due to the fitting with a power function is displayed.
 The horizontal lines represent the values predicted from the spin-wave approximation. }
 \label{fig:correlation}
\end{figure}

We first calculate the spin correlation function $C(r)$ for the XY model ($N=2$).
We start from a random initial condition and obtain a steady state after a sufficiently long time.
The correlation function is calculated from a field configuration $\bvec{\phi}(\bvec{r})$ of the steady state.
For a sufficiently large system, the average with respect to the random field and the thermal noise can be replaced by 
the spatial average because of the self-averaging property.
We also take the time average, which is equivalent to the ensemble average with respect to the random field and the thermal noise.
The detailed explanation for the method of the numerical calculation of the correlation function is presented in the Appendix B.
The correlation functions for different values of temperature are displayed in Fig.~\ref{fig:correlation}.
The upper (a) and lower (b) panels depict $C(r)$ for the directions parallel and perpendicular to the driving velocity $\bvec{v}$, respectively.
In order to verify the finite size effect, we calculate $C(r)$ for system sizes of $60^3$, $100^3$, and $150^3$.
The system size dependence of $C(r)$ for the perpendicular direction is larger than that for the parallel direction because of the periodic boundary conditions.
We can observe the phase transition from the low temperature regime, in which the correlation function exhibits a power-law decay, 
to the high temperature regime, in which it displays an exponential decay.
The panel (c) represents the exponents as a function of temperature.
The horizontal lines represent the theoretical prediction of the spin-wave approximation.
The exponents $\alpha_{\parallel}$ and $\alpha_{\perp}$ increase linearly at low temperatures, but they exhibit strong temperature dependence near the transition temperature.

\begin{figure}
 \centering
 \includegraphics[width=80mm]{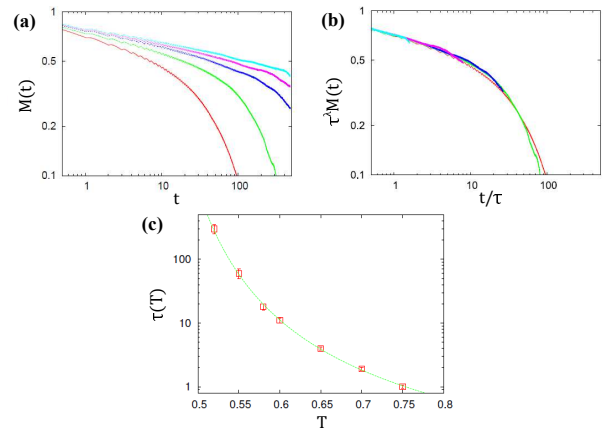}
 \caption{(Color online) Panel (a): Relaxation of the magnetization $M(t)$ for different values of temperature. The other parameters are $h_0=1.4$ and $v=1.0$.
 The values of temperature are $T=0.52,\: 0.55,\: 0.58,\: 0.65$, and $0.75$ from the up to the bottom.
 Panel (b): Scaling plot of the magnetization to Eq.~(\ref{scaling}) with appropriately chosen $\tau(T)$ and $\lambda$.
 $ \lambda=0.08 $ and $ \tau(0.75)=1,\: \tau(0.65)=4.0,\: \tau(0.58)=18,\: \tau(0.55)=60,\: \tau(0.52)=300. $
 Panel (c): Relaxation time $\tau$ as a function of temperature.
 The curve fitted to Eq.~(\ref{rel-time}) with $T_{\rm c}=0.43$ is shown. }
 \label{fig:magnetization}
\end{figure}

We next determine the transition temperature $T_{\rm c}$ as a function of the strength of the random field $h_0$ 
and the driving velocity $v$ by using the non-equilibrium relaxation method \cite{Ito07}.
With this method, we observe the relaxation of the magnetization $\bvec{M}(t)=V^{-1} \int \bvec{\phi}(\bvec{r}) {\rm d} \bvec{r}$ 
from the complete ordered state $\phi^1(\bvec{r}) \equiv 1$ and $\phi^2(\bvec{r}) \equiv 0 $.
The asymptotic behavior of the magnetization for $T \geq T_{\rm c}$ is summarized as
\begin{eqnarray}
M(t) \sim
  \begin{cases}
    {\rm exp}(-t/\tau(T)), \:\:\:\:\:(T > T_{\rm c}), & \\
    t^{-\lambda}, \:\:\:\:\:(T = T_{\rm c}), &
  \end{cases}
\end{eqnarray}
where $\tau(T)$ is the relaxation time.
In order to determine the relaxation time as a function of temperature for the disordered phase,
we assume the following scaling form,
\begin{equation}
M(t)=\tau(T)^{-\lambda}m(t/\tau(T)),
\label{scaling}
\end{equation}
for $T > T_{\rm c}$.
The relaxation of the magnetization $M(t)$ and its scaling plot are displayed in the panels (a) and (b) of Fig.~\ref{fig:magnetization}.
The system size is $60^3$.
100 independent runs are performed for averaging.
From the analogy to the KT transition of the two-dimensional pure XY model \cite{Ito03}, 
the correlation length is expected to diverge exponentially as $\xi \sim {\rm exp}(a/\sqrt{T-T_{\rm c}})$.
Thus, we assume that the relaxation time diverges in the same way,
\begin{equation}
\tau(T) = B \: {\rm exp} \left( \frac{A}{\sqrt{T-T_{\rm c}}} \right).
\label{rel-time}
\end{equation}
Fitting $\tau(T)$ into Eq.~(\ref{rel-time}) with parameters $A$, $B$, and $T_{\rm c}$, we obtain the transition temperature.
The best fitting is shown in the panel (c) of Fig.~\ref{fig:magnetization}.

\begin{figure}
 \centering
 \includegraphics[width=80mm]{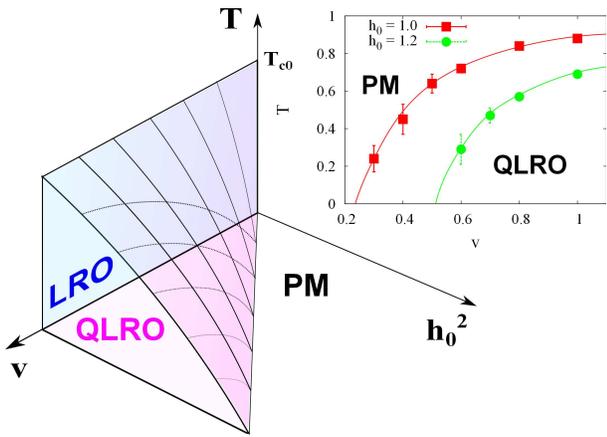}
 \caption{(Color online) Schematic phase diagram with respect to the disorder $h_0$, driving velocity $v$, and temperature $T$.
 The parameter $T_{\rm c0}$ denotes the transition temperature of the three-dimensional pure XY model.
 The abbreviation PM denotes the paramagnetic or disordered phase.
 The inset depicts $T_{\rm c}$ as a function of $v$.
 The solid lines serve as a visual guide.
 The squares (red) and circles (green) denote the values of $T_{\rm c}$ for $h_0=1.0$ and $h_0=1.2$, respectively. }
 \label{fig:phase}
\end{figure}

Fig.~\ref{fig:phase} displays the schematic phase diagram with respect to the strength of the disorder $h_0$, driving velocity $v$, and temperature $T$.
The QLRO phase appears in the large-$v$ and low-$T$ regime.
In the region in which $h_0=0$, the LRO phase exists because the model is identical to the three-dimensional pure XY model in the moving frame.
The small-$v$ and high-$T$ regime corresponds to the disordered phase.
The phase boundary at $T=0$ is given by $\Gamma h_0^2 \sim Kv$.
It is noteworthy that the infinitesimally small random field breaks the LRO and leads to the QLRO.
For an arbitrarily large value of $h_0$, the QLRO is observed for sufficiently large values of $v$.
We have checked that Eq.~(\ref{scale-separation}) is satisfied in the region of the QLRO phase displayed in the phase diagram Fig.~\ref{fig:phase} at low temperatures.
This justifies the validity of the spin-wave approximation.

\begin{figure}
 \centering
 \includegraphics[width=60mm]{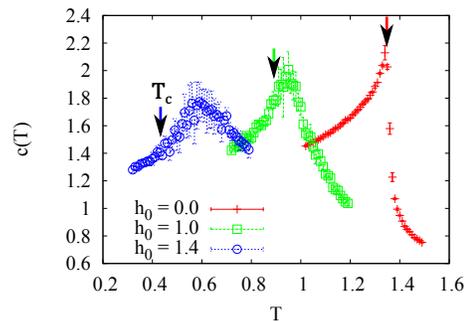}
 \caption{(Color online) Specific heat $c(T)$ as a function of temperature for system size of $60^3$.
 The driving velocity is $v=1.0$.
 The crosses (red), squares (green), and circles (blue) denote $c(T)$ values for $h_0=0.0, \:1.0$, and $1.4$, respectively.
 The upper arrows represent the transition temperature $T_{\rm c}$ as determined by the non-equilibrium relaxation method. }
 \label{fig:specific-heat}
\end{figure}

In order to compare this transition with the KT transition of the two-dimensional pure XY model, we calculate the specific heat as a function of temperature.
We define the specific heat as $c(T)=\partial \langle H \rangle_{\rm ss}/\partial T$, where $ \langle ...\rangle_{\rm ss}$ represents the average with respect to the non-equilibrium steady state.
Since the system is not in equilibrium, this specific heat is not related to the energy fluctuation.
Fig.~\ref{fig:specific-heat} shows the specific heat as a function of temperature for different values of the disorder strength $h_0$.
The upper arrows represent the transition temperature $T_{\rm c}$ as determined by the non-equilibrium relaxation method. 
The case that $h_0=0$ corresponds to the three-dimensional pure XY model.
In the presence of a finite amount of disorder, the discontinuity of $c(T)$ disappears and a smooth peak remains above the transition temperature.
The position of the peak decreases with increase in the strength of the disorder.
It is to be noted that the specific heat does not exhibit any singularity at the transition temperature.
The absence of a singularity and the existence of the smooth peak resemble the KT transition of the two-dimensional pure XY model \cite{Gupta}.

\subsection{Heisenberg model}

\begin{figure}
 \centering
 \includegraphics[width=55mm]{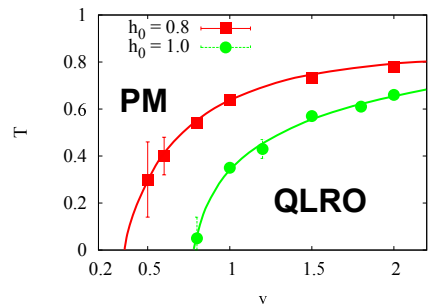}
 \caption{(Color online) Transition temperature $T_{\rm c}$ as a function of $v$ for the Heisenberg model ($N=3$).
 The squares (red) and circles (green) denote $T_{\rm c}$ values for $h_0=0.8$ and $h_0=1.0$, respectively. 
 The solid lines serve as a visual guide.}
 \label{fig:phase-Heisenberg}
\end{figure}

\begin{figure}
 \centering
 \includegraphics[width=60mm]{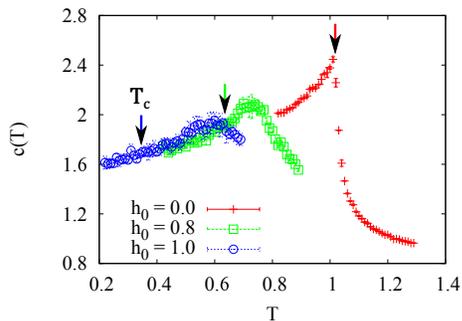}
 \caption{(Color online) Specific heat $c(T)$ as a function of temperature for the Heisenberg model ($N=3$).
 The driving velocity is $v=1.0$.
 The crosses (red), squares (green), and circles (blue) denote $c(T)$ values for $h_0=0.0, \:0.8$, and $1.0$, respectively.
 The upper arrows represent the transition temperature $T_{\rm c}$ as determined by the non-equilibrium relaxation method. }
 \label{fig:specific-heat-Heisenberg}
\end{figure}

Next, we consider the Heisenberg model ($N=3$).
We found that the phase diagram of the Heisenberg model is qualitatively similar to that of the XY model.
Fig.~\ref{fig:phase-Heisenberg} shows the transition temperature determined by using the non-equilibrium relaxation method.
We display the specific heat as a function of temperature in Fig.~\ref{fig:specific-heat-Heisenberg}.
In the presence of a finite amount of disorder, the discontinuity of $c(T)$ disappears and a smooth peak remains above the transition temperature.
The presence of the QLRO for $N=3$ contrasts with the cases of the two-dimensional pure O($N$) models, 
in which the KT transition does not exist for $N=3$ \cite{Zinn-Justin}.

\section{Conclusions}
\label{sec:conclusions}
We investigated the non-equilibrium phase transition of three-dimensional O($N$) models driven with a uniform velocity over a quenched random field.
For the cases that $N=2$ and $3$ (XY and Heisenberg model, respectively), we show that QLRO appears in the strong driving regime.
Furthermore, the transition from the QLRO phase to the disordered phase resembles the KT transition in the two-dimensional pure XY model.
It is noteworthy that there exists a critical value $N_{\rm c}$ such that QLRO does not exist for $N > N_{\rm c}$.
The renormalization group analysis is required to determine this value of $N_{\rm c}$.
Moreover, it is also necessary to clarify the relation between this QRLO and the Bragg glass phase in the vortex lattice systems \cite{LeDoussal94,LeDoussal98}.
We plan to investigate these theoretical aspects in our future studies.

Finally, we remark on the topics related to the experimental realization of the QLRO under consideration.
We consider nematic liquid crystals (NLCs) flowing in a random substrate such as an aerogel or a porous medium.
Recently, the dynamics of liquid crystals confined to a complex geometry has attracted considerable attention due to not only fundamental research interest but also its industrial applications \cite{Araki,Sengupta}.
For NLCs in a random substrate, the random anchoring, which results from the complicated surface structure, significantly influences the ordering structure and thermodynamic properties \cite{Marinelli,Petridis}.
Here, we remark that certain factors concerning NLCs flowing in the random substrate are not included in our models.
For example, we ignore the inhomogeneity of the velocity field.
The inhomogeneity of the velocity field acts as an additional random perturbation for the directors of the NLCs.
Since the correlation of the spatial fluctuation of hydrodynamic velocity field exhibits power-law decay, 
this random perturbation is qualitatively different from the random anchoring whose correlation decays over a short distance.
The consequences of this hydrodynamic effect and other factors excluded in our models should be investigated in future work.

\begin{acknowledgments}
The author thanks Shin-ich Sasa for fruitful discussions.
The present study was supported by JSPS KAKENHI No.15J01614 and by the JSPS Core-to-Core program
``Non-equilibrium dynamics of soft-matter and information.''
\end{acknowledgments}

\setcounter{equation}{0}
\def\theequation{A\arabic{equation}}

\appendix
\section{Calculation of the spin-wave approximation}
We investigate the asymptotic behavior of the mean square relative displacement Eq.~(\ref{MSRD}).
First, we consider the case in which $\bvec{r_1}-\bvec{r_2}$ is parallel to $\bvec{v}$.
Eq.~(\ref{MSRD}) is then rewritten in terms of the polar coordinate
\begin{eqnarray}
\langle (\theta(\bvec{r_1})-\theta(\bvec{r_2}))^2 \rangle \nonumber \\
= 2 \Gamma^2 h_0^2 \int^{\infty}_{0} \frac{{\rm d} q}{(2 \pi)^3} \int^{2 \pi}_{0} {\rm d} \phi \int^{\pi}_{0} {\rm d} \theta q^2 {\rm sin} \theta \nonumber \\
\times \frac{1-{\rm cos} \left\{ q  |\bvec{r_1}-\bvec{r_2}|  {\rm cos} \theta \right\} }{\Gamma^2 K^2 q^4+v^2 q^2 {\rm cos}^2 \theta} \nonumber \\
= \frac{\Gamma^2 h_0^2}{2 \pi^2} \int^{\infty}_{0} {\rm d} q \int^{1}_{-1} {\rm d} u \frac{1-{\rm cos} \left\{ q |\bvec{r_1}-\bvec{r_2}| u \right\} }{\Gamma^2 K^2 q^2+v^2 u^2 },
\end{eqnarray}
where we have changed the variable $ u={\rm cos}\theta $.
Since we are interested in the contribution from the small wave number regime $q \ll v/(\Gamma K)$, the range of the $u$-integral can be extended to $(-\infty,\: \infty)$.
Thus, we have
\begin{eqnarray}
\langle (\theta(\bvec{r_1})-\theta(\bvec{r_2}))^2 \rangle \simeq \frac{\Gamma h_0^2}{2 \pi Kv} \nonumber \\
\times \int^{v/(\Gamma K)}_{0} {\rm d} q \frac{1}{q} \left\{1-{\rm exp} \left(-\frac{\Gamma Kq^2}{v} |\bvec{r_1}-\bvec{r_2}| \right) \right\}.
\end{eqnarray}
The integrand takes the maximal value at $q=q_{\rm m} \sim \sqrt{v/(\Gamma K |\bvec{r_1}-\bvec{r_2}|)}$, and it decays as $q^{-1}$ for $q \gg q_{\rm m}$.
Therefore, we obtain
\begin{eqnarray}
\langle (\theta(\bvec{r_1})-\theta(\bvec{r_2}))^2 \rangle \nonumber \\
\simeq \frac{\Gamma h_0^2}{2 \pi Kv} \int^{v/(\Gamma K)}_{q_{\rm m}} \frac{{\rm d} q}{q} + {\rm const} \nonumber \\
= \frac{\Gamma h_0^2}{2 \pi Kv} {\rm log} \frac{v/(\Gamma K)}{q_{\rm m}} + {\rm const} \nonumber \\
= \frac{\Gamma h_0^2}{4 \pi Kv} {\rm log} |\bvec{r_1}-\bvec{r_2}| + {\rm const}.
\end{eqnarray}
This is one of Eqs.~(\ref{MSRD-asymp}).

We next consider the case in which $\bvec{r_1}-\bvec{r_2}$ is perpendicular to $\bvec{v}$.
Eq.~(\ref{MSRD}) is then rewritten as
\begin{eqnarray}
\langle (\theta(\bvec{r_1})-\theta(\bvec{r_2}))^2 \rangle \nonumber \\
=2 \Gamma^2 h_0^2 \int \frac{{\rm d}^3 \bvec{q}}{(2 \pi)^3} \frac{1-{\rm cos} \left\{ q_z |\bvec{r_1}-\bvec{r_2}| \right\} }{\Gamma^2 K^2 q^4+v^2 q_x^2},
\end{eqnarray}
where $\bvec{r_1}-\bvec{r_2}$ is assumed to be parallel to the $z$-axis.
Since we are interested in the contribution from the small wave number regime $q \ll v/(\Gamma K)$, the term containing $q_x$ in $q^4$ of the denominator can be ignored.
Thus, we have
\begin{eqnarray}
\langle (\theta(\bvec{r_1})-\theta(\bvec{r_2}))^2 \rangle \nonumber \\
\simeq 2 \Gamma^2 h_0^2 \int_{q < v/(\Gamma K)} \frac{{\rm d}^3 \bvec{q}}{(2 \pi)^3} \frac{1-{\rm cos} \left\{ q_z |\bvec{r_1}-\bvec{r_2}| \right\} }{\Gamma^2 K^2 (q_y^2+q_z^2)^2+v^2 q_x^2}, \nonumber \\
= \frac{\Gamma h_0^2}{ \pi Kv} \int_{q < v/(\Gamma K)} \frac{{\rm d} q_y}{2 \pi} \frac{{\rm d} q_z}{2 \pi} \frac{1-{\rm cos} \left\{ q_z |\bvec{r_1}-\bvec{r_2}| \right\} }{q_y^2+q_z^2} \nonumber \\
= \frac{\Gamma h_0^2}{ \pi Kv} \int^{v/(\Gamma K)}_{0} \frac{{\rm d} q}{2 \pi} \frac{1-J_0(q |\bvec{r_1}-\bvec{r_2}|)}{q} \nonumber \\
\simeq \frac{\Gamma h_0^2}{2 \pi Kv} {\rm log} |\bvec{r_1}-\bvec{r_2}| + {\rm const},
\end{eqnarray}
where $J_0(x)$ is the Bessel function. This is the other one of Eqs.~(\ref{MSRD-asymp}).

\setcounter{equation}{0}
\def\theequation{B\arabic{equation}}

\section{Numerical calculation of the correlation function}
We explain the method of the numerical calculation of the correlation function.
The correlation function is defined as
\begin{eqnarray}
C(\bvec{r}) = \int D \bvec{h} \: P_{\rm RF}[\bvec{h}] \int D \bvec{\phi} \: \bvec{\phi}(\bvec{r}) \cdot \bvec{\phi}(\bvec{0}) P_{\rm st}[\bvec{\phi};\bvec{h}],
\label{corr-def}
\end{eqnarray}
where $P_{\rm st}[\bvec{\phi};\bvec{h}]$ denotes the steady state solution of the Fokker-Planck equation corresponding to Eq.~(\ref{EM-LF})
for a fixed random field $ \bvec{h}(\bvec{r}) $ and $P_{\rm RF}[\bvec{h}]$ is the distribution function for $ \bvec{h}(\bvec{r}) $.

We implement the numerical calculation in a moving frame with $\bvec{v}$.
Thus, we define $\bvec{r}' = \bvec{r}-\bvec{v}t$ and $ \bvec{\phi}'(\bvec{r}') = \bvec{\phi}(\bvec{r}) $.
Eq.~(\ref{EM-LF}) is then rewritten as follows:
\begin{equation}
\frac{\partial \bvec{\phi}'(\bvec{r}',t)}{\partial t}=-\Gamma \frac{\delta H[\bvec{\phi}';\bvec{h}_t]}{\delta \bvec{\phi}'(\bvec{r}',t)}+\bvec{\eta}(\bvec{r}',t),
\label{EM-CM}
\end{equation}
where $\bvec{h}_t(\bvec{r}') \equiv \bvec{h}(\bvec{r}'+\bvec{v}t)$ is a moving random field.

The numerical calculation is implemented in a cubic simulation box, whose size is $L$.
We numerically solve Eq.~(\ref{EM-CM}) with the periodic and free boundary conditions 
for the directions perpendicular and parallel to the driving velocity $\bvec{v}$, respectively.
In what follows, we omit the prime symbols from Eq.~(\ref{EM-CM}).
The random field is continuously generated in the front boundary of the simulation box $x=L$, and it moves with velocity $-\bvec{v}$.
$\bvec{h}_{\rm B}(\bvec{r}_{\perp},t)$ denotes the random field generated at the front boundary, 
where $\bvec{r}_{\perp}=(y,z)$ represents a coordinate of the front boundary.
Then, the moving random field is written as $\bvec{h}_t(x,y,z)=\bvec{h}_{\rm B}(y,z,t-(L-x)/v)$.
We denote the time dependent solution of the Fokker-Planck equation corresponding to Eq.~(\ref{EM-CM})
for a realization of $\bvec{h}_{\rm B}(\bvec{r}_{\perp},s)$ by $P^{\rm CM}_t[\bvec{\phi}; \{ \bvec{h}_{\rm B}(s) \}_{s \leq t}]$.
Note that $P^{\rm CM}_t$ depends on all history of the random field $\bvec{h}_{\rm B}$.
From the equivalence of Eqs.~(\ref{EM-LF}) and (\ref{EM-CM}), Eq.~(\ref{corr-def}) is rewritten as follows:
\begin{eqnarray}
C(\bvec{r}) = \int D \bvec{h}_{\rm B} \: P_{\rm RF}[\bvec{h}_{\rm B}] \nonumber \\
\times \int D \bvec{\phi} \: \bvec{\phi}(\bvec{r}) \cdot \bvec{\phi}(\bvec{0}) P^{\rm CM}_t[\bvec{\phi}; \{ \bvec{h}_{\rm B}(s) \}_{s \leq t}],
\label{corr-CM}
\end{eqnarray}
for sufficiently large $t$, where $P_{\rm RF}[\bvec{h}_{\rm B}]$ is the distribution function for the random field generated at the front boundary.

If the simulation box is sufficiently large $L \gg \xi$, where $ \xi $ is the correlation length, we can assume the self-averaging property.
Thus, the average with respect to the random field and the thermal noise can be replaced by the spatial average
\begin{equation}
C(\bvec{r}) = \frac{1}{V} \int d \bvec{r}' \: \bvec{\phi}(\bvec{r}'+\bvec{r}) \cdot \bvec{\phi}(\bvec{r}'),
\label{corr-by-self-average}
\end{equation}
where $ \bvec{\phi}(\bvec{r}) $ is a solution of the Langevin equation (\ref{EM-CM}).
We checked that the dependence of the correlation functions Eq.~(\ref{corr-by-self-average}) on realizations of
$\bvec{h}_{\rm B}(\bvec{r}_{\perp},t)$ and $ \bvec{\eta}(\bvec{r},t) $ is small.
We also take the time average of Eq.~(\ref{corr-by-self-average}) at time intervals of $L/v$.


\end{document}